\documentstyle[twoside,fleqn,espcrc2]{article}

\newcommand{\AmS}{{\protect\the\textfont2
  A\kern-.1667em\lower.5ex\hbox{M}\kern-.125emS}}

\def\slashchar#1{\setbox0=\hbox{$#1$}           
   \dimen0=\wd0                                 
   \setbox1=\hbox{/} \dimen1=\wd1               
   \ifdim\dimen0>\dimen1                        
      \rlap{\hbox to \dimen0{\hfil/\hfil}}      
      #1                                        
   \else                                        
      \rlap{\hbox to \dimen1{\hfil$#1$\hfil}}   
      /                                         
   \fi}                                         %
 
\def \half {{\scriptstyle {1\over 2}}}

\input epsf
\input colordvi  

\title{Is the doubler of the electron an antiquark?}

\author{Michael Creutz
\address{Physics Department, Brookhaven
 National Laboratory, PO Box 5000, Upton, NY 11973-5000, USA\\
 creutz@wind.phy.bnl.gov
}
\thanks{This manuscript has been authored under contract number
DE-AC02-76CH00016 with the U.S.~Department of Energy.  Accordingly,
the U.S.~Government retains a non-exclusive, royalty-free license to
publish or reproduce the published form of this contribution, or allow
others to do so, for U.S.~Government purposes.}
}
 
\begin{document}

\begin{abstract}
Reassigning degrees of freedom in the standard model allows an
interpretation where the SU(2) gauge group is vector-like and parity
violation moves to the strong and electromagnetic interactions.  In
this picture, the electron is paired with an anti-quark.  Requiring
exact gauge invariance for the electromagnetic interaction clarifies
the mechanism behind a recent proposal for a lattice regularization of
the standard model.
\end{abstract}

\maketitle

The standard model of elementary particle interactions is based on the
product of three gauge groups, $SU(3)\times SU(2) \times U(1)_{em}$.
Here the $SU(3)$ represents the strong interactions of quarks and
gluons, the $U(1)_{em}$ corresponds to electromagnetism, and the
$SU(2)$ gives rise to the weak interactions.  I ignore here the
technical details of electroweak mixing.

The full model is, of course, parity violating, as necessary to
describe observed helicities in beta decay.  This violation is
normally considered to lie in the $SU(2)$ of the weak interactions,
with both the $SU(3)$ and $U(1)_{em}$ being parity conserving.
However, this is actually a convention, adopted primarily because the
weak interactions are small.  I argue below that reassigning degrees
of freedom allows a reinterpretation where the $SU(2)$ gauge
interaction is vector-like.  Since the full model is parity violating,
this process shifts the parity violation into the strong,
electromagnetic, and Higgs interactions.  The resulting theory pairs
the left handed electron with a right handed anti-quark to form a
Dirac fermion.

With a vector-like weak interaction, the chiral issues which complicate
lattice formulations now move to the other gauge groups.  Requiring
gauge invariance for the re-expressed electromagnetism then clarifies
the mechanism behind a recent proposal for a lattice regularization of
the standard model.

To begin, consider only the first generation, which involves four left
handed doublets.  These correspond to the neutrino/electron lepton
pair plus three colors for the up/down quarks
\begin{equation}
\pmatrix{\nu \cr e^-\cr}_L,
\ \pmatrix{{u^r} \cr {d^r}\cr}_L,
\ \pmatrix{{u^g} \cr {d^g}\cr}_L,
\ \pmatrix{{u^b} \cr {d^b}\cr}_L
\end{equation}
Here the superscripts from the set $\{r,g,b\}$ represent the internal
$SU(3)$ index of the strong interactions, and the subscript $L$ indicates
left-handed helicities. 

If I ignore the strong and electromagnetic interactions, leaving only
the weak $SU(2)$, each of these four doublets is
equivalent and independent.  I now arbitrarily pick two of them
and do a charge conjugation operation, thus switching to their
antiparticles
\begin{equation}\matrix{
\pmatrix{{u^g} \cr {d^g}\cr}_L \longrightarrow 
\pmatrix{\overline{{d^g}} \cr \overline{{u^g}}\cr}_R \cr
\pmatrix{{u^b} \cr {d^b}\cr}_L \longrightarrow 
\pmatrix{\overline{{d^b}} \cr \overline{{u^b}}\cr}_R \cr
}
\end{equation}
In four dimensions anti-fermions have the opposite helicity; so, I
label these new doublets with $R$ representing right handedness.

With two left and two right handed doublets, I now combine them into
two Dirac doublets
\begin{equation}
\pmatrix{
\pmatrix{\nu \cr e^-\cr}_L\cr
\pmatrix{\overline{{d^g}} \cr \overline{{u^g}}\cr}_R\cr
}
\qquad
\pmatrix{
\pmatrix{{u^r} \cr {d^r}\cr}_L\cr
\pmatrix{\overline{{d^b}} \cr \overline{{u^b}}\cr}_R \cr
}
\end{equation}
Formally in terms of the underlying fields, the construction takes
\begin{equation}\matrix{
\psi=\half (1-\gamma_5)\psi_{(\nu,e^-)}+\half (1+\gamma_5)
\psi_{({\overline{d^g}},{\overline{u^g}})} \cr
\chi=\half (1-\gamma_5)\psi_{({u^r}, {d^r})}+\half (1+\gamma_5)
\psi_{({\overline{d^b}},{\overline{u^b}})} \cr
}
\end{equation}

From the conventional point of view these fields have rather peculiar
quantum numbers.  For example, the left and right parts have different
electric charges.  Electromagnetism now violates parity.  The left and
right parts also have different strong quantum numbers; the strong
interactions violate parity as well.  Finally, the components have
different masses; parity is violated in the Higgs mechanism.

The different helicities of these fields also have variant baryon
number.  This is directly related to the known baryon violating
processes through weak ``instantons'' and axial
anomalies\cite{thooft}.  When a topologically non-trivial weak field
is present, the axial anomaly arises from a level flow out of the
Dirac sea \cite {anomalyflow}.  This generates a spin flip in the
fields, {\it i.e.} $e^-_L \rightarrow ({\overline{u^g}})_R$.  Because
of the peculiar particle identification, this process does not
conserve charge, with $\Delta Q= -{2\over 3} +1={1\over 3}$.  This
would be a disaster for electromagnetism were it not for the fact that
simultaneously the other Dirac doublet also flips {${d^r}_L
\rightarrow ({\overline{u^b}})_R$} with a compensating $\Delta Q =
-{1\over 3}$.  This is anomaly cancelation, with the total $\Delta Q
= {1\over 3}-{1\over 3}=0$.  Only when both doublets are considered
together is the $U(1)$ symmetry restored.  In this anomalous process
baryon number is violated, with $L+Q\rightarrow \overline Q +
\overline Q$.  This is the famous `` `t Hooft vertex'' \cite {thooft}.

So far the discussion has been in the continuum.  Now I turn to the
lattice, and use the Kaplan-Shamir approach for fermions
\cite{kaplanshamir}.  In this picture, our four dimensional world is a
``4-brane'' embedded in 5-dimensions.  The complete lattice is a five
dimensional box with open boundaries, and the parameters are chosen so
the physical quarks and leptons appear as surface zero modes.  The
elegance of this scheme lies in the natural chirality of these modes
as the size of the extra dimension grows.

With a finite fifth dimension a doubling phenomenon remains, coming
from interfaces appearing as surface/anti-surface pairs.  It is
natural to couple a four dimensional gauge field equally to both
surfaces, giving rise to a vector-like theory.

\begin{figure}
\epsfxsize .9 \hsize
\centerline{\epsffile{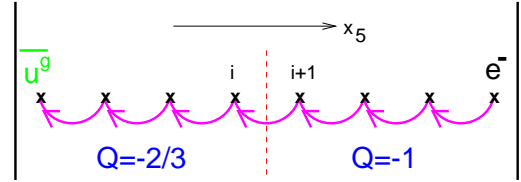}}
\caption{Pairing the electron with the anti-green-up-quark.}
\label{fig:1}
\end{figure}

I now insert the above pairing into this five dimensional scheme.  In
particular, I consider the left handed electron as a zero mode on one
wall and the right handed anti-green-up-quark as the partner zero mode
on the other wall, as sketched in Fig.~\ref{fig:1}.  This provides a
lattice regularization for the $SU(2)$ of the weak interactions.

However, since these two particles have different electric charge,
$U(1)_{EM}$ must be broken in the interior of the extra dimension.  I
now proceed in analogy to the ``waveguide'' picture\cite{waveguide}
and restrict this charge violation to $\Delta Q$ to one layer at some
interior $x_5=i$.  Then the fermion hopping term from $x_5=i$ to $i+1$
\begin{equation}
\overline\psi_{i}P\psi_{i+1}\qquad{(P=\gamma_5+r)}
\end{equation}
is a $Q=1/3$ operator.  At this layer, electric charge is not
conserved.  This is unacceptable and needs to be fixed.

\begin{figure}
\epsfxsize .7 \hsize
\centerline{\epsffile{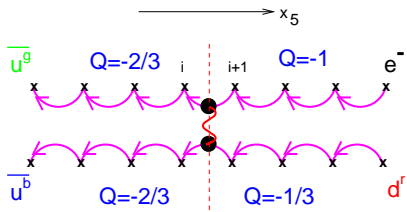}}
\caption {Transferring charge between the doublets.}
\label{fig:2}
\end{figure}

To restore the $U(1)$ symmetry one must transfer the charge from
$\psi$ to the compensating doublet $\chi$.  For this I replace
the sum of hoppings with a product on the offending layer
\begin{equation}
\overline\psi_{i}P\psi_{i+1}
{+}\overline\chi_{i}P\chi_{i+1}
\Black{\longrightarrow}
\overline\psi_{i}P\psi_{i+1}
{\times}\overline\chi_{i}P\chi_{i+1}
\end{equation}
This introduces an electrically neutral four fermi operator.  Note
that it is baryon violating, involving a ``lepto-quark/diquark''
exchange, as sketched in Fig.~\ref{fig:2}.  One might think of the
operator as representing a ``filter'' at $x_5=i$ through which only
charge compensating pairs of fermions can pass.

In five dimensions there is no chiral symmetry.  Even for the free
theory, combinations like $\overline\psi_{i}P\psi_{i+1} $ have vacuum
expectation values.  I use such as a ``tadpole,'' with $\chi$
generating an effective hopping for $\psi$ and {\it vice versa}.

Actually the above four fermion operator is not quite sufficient for
all chiral anomalies, which can also involve right handed
singlet fermions.  To correct this I need explicitly include the right
handed sector, adding similar four fermion couplings
(also electrically neutral).

Having fixed the $U(1)$ of electromagnetism, I restore the strong
$SU(3)$ with an antisymmetrization $ {Q^r}{Q^g}{Q^b}{\longrightarrow
\epsilon^{\alpha\beta\gamma}Q^\alpha Q^\beta Q^\gamma}$.  Note that
similar left-right inter-sector couplings are needed to correctly
obtain the effects of topologically non-trivial strong gauge fields.

For the strong interactions alone I could replace the four fermion
vertices with simple hoppings, matching particles with their usual
partners.  This leads to the usual chiral predictions, such as
$M_Q\sim m_\pi^2$ in $L_5\rightarrow \infty$ limit, and forms the
basis of the formulation of Ref.~\cite{furmanshamir}.  Recent tests
are encouraging that this may be numerically advantageous
\cite{blumsoni}.

An alternative view is to fold the lattice about the interior of the
fifth dimension, placing all light modes on one wall and having the
multi-fermion operator on the other.  This is the model of Ref.~\cite
{us}, with the additional inter-sector couplings correcting a
technical error \cite{neuberger}.

Unfortunately the scheme is still non rigorous.  In particular, the
non-trivial four fermion coupling represents a new defect and we need
to show that this does not give rise to unwanted extra zero modes.  Note,
however, that the five dimensional mass is the same on both sides of
defect, removing topological reasons for such.

A second worry is that the four fermion coupling might induce an
unwanted spontaneous symmetry breaking of one of the gauge symmetries.
We need a strongly coupled paramagnetic phase without spontaneous
symmetry breaking.  Ref.~\cite{us} showed that strongly coupled zero
modes preserved the desired symmetries, but the analysis ignored
contributions from heavy modes in the fifth dimension.

Assuming all works as desired, the model raises several other
interesting questions.  As formulated, I used a right handed neutrino
to provide all quarks with partners.  Is there some variation that
avoids this particle, which completely decouples in the continuum
limit?  Another question concerns possible numerical simulations; is
the effective action positive?  Finally, we have used the details of
the usual standard model, leaving open the question of whether this
model is somehow special.  Can we always use multi-fermion couplings
to eliminate undesired modes in other anomaly free chiral theories?
There is much more to do!

\end{document}